# Route-Based Detection of Conflicting ATC Clearances on Airports

Benjamin Weiß, Federico Centarti, Felix Schmitt, Stephen Straub

DFS Deutsche Flugsicherung GmbH
SH/P Systemhaus, Am DFS-Campus 7, 63225 Langen, Germany
benjamin.weiss@dfs.de, federico.centarti@dfs.de, felix.schmitt@dfs.de, stephen.straub@dfs.de

*Abstract* — Runway incursions are among the most serious safety concerns in air traffic control. Traditional A-SMGCS level 2 safety systems detect runway incursions with the help of surveillance information only. In the context of SESAR, complementary safety systems are emerging that also use other information in addition to surveillance, and that aim at warning about potential runway incursions at earlier points in time. One such system is "conflicting ATC clearances", which processes the clearances entered by the air traffic controller into an electronic flight strips system and cross-checks them for potentially dangerous inconsistencies. The cross-checking logic may be implemented directly based on the clearances and on surveillance data, but this is cumbersome. We present an approach that instead uses ground routes as an intermediate layer, thereby simplifying the core safety logic.

## I. Introduction

Preventing *runway incursions* is an important safety goal in air traffic control (ATC). As defined by ICAO [1], a runway incursion is "any occurrence at an aerodrome involving the incorrect presence of an aircraft, vehicle or person on the protected area of a surface designated for the landing and take-off of aircraft". We understand this to include all occurrences where the required *runway separation* between two mobiles (i.e., aircraft or vehicles) is lost, as well as all occurrences where a mobile is present on a runway without ATC permission (even if runway separation is maintained). The rules for runway separation are defined by ICAO [2] and local implementations thereof. For example, ICAO [2] demands – roughly speaking – that at the moment when a departing aircraft commences take-off or when a landing aircraft overflies the runway threshold, there must be no traffic on the runway in front of it. If such a rule is violated, then this constitutes a runway incursion.

Besides other safety provisions, one line of attack against runway incursions is to use automated systems that detect runway incursions or their precursors, and that warn air traffic controllers (ATCOs) and/or pilots and vehicle drivers about such situations [3],[4]. Traditionally, such systems are based exclusively on surveillance information: they continuously monitor the positions and movement vectors of mobiles on and around runways, and emit a warning to the ATCO if a dangerous situation is unfolding. Such systems are known as *runway incursion monitoring systems (RIMS)*, or as *advanced surface movement guidance and control system (A-SMGCS) level 2 alerting* systems [5],[6],[7].

RIMS can provide reliable warnings only on short notice, requiring an immediate ATCO reaction. In the context of the *Single European Sky ATM Research Programme (SESAR)* – and pioneered by EUROCONTROL's *Integrated Tower Working Position (ITWP)* project [8] – two new safety systems complementary to RIMS are emerging, which aim at warning about potential runway incursions at earlier points in time. The key to both systems, called *conflicting ATC clearances (CATC)* [9] and *conformance monitoring* [10], is that in addition to surveillance information, they also make use of the *clearances* entered by ATCOs into the system. In other words, they are made possible by the interaction of two functional areas that are traditionally allocated to separate systems, namely a *sensor data processing system* dealing with geographical positions of mobiles on the ground and in the air, and an *electronic flight strips system* dealing with flight plans and clearances.

Where conformance monitoring is about checking whether a mobile's behaviour conforms to its clearance, CATC is about cross-checking clearances for consistency. For example, a CATC system can detect if an aircraft A receives a "line up" clearance while another aircraft B holds a "land" clearance for the same runway, unless B has already passed the runway entry to be used by A, in which case no warning is given. The system thus guards against rare but potentially dangerous ATCO mistakes, where clearances are issued that, in combination, could lead to a violation of runway separation. A recent study suggests that as many as half of all European runway incursions may have such causes [11].

CATC safety logic can be implemented by directly making use of surveillance and clearance information, but this involves lots of special case handling: is the runway entry of A in front of or behind the current position of B? What if there are intersecting runways? We describe an approach to CATC that is instead based on *ground routing*, i.e., on a separate system functionality that is concerned with the routes that mobiles take on the airport's network of runways, taxiways, apron taxilanes and parking stands. The availability of ground routes simplifies the CATC problem significantly: clearances can be mapped to routes by distinguishing between a cleared and a not-yet-cleared route part, and the core safety logic can then work with thus enhanced routes instead of directly with clearances and surveillance data. Essentially, it has to check for overlaps between cleared route segments on runways.

*Outline.* In Section II, the concept of CATC is described in more detail. We provide background on ground routing in Section III, before presenting the route-based approach to CATC in Section IV. Section V describes how conditional clearances can be handled. Section VII is about the presentation of warnings in a human machine interface, and Section VII gives an outlook on a possible future extension to trajectory-based CATC. Section VIII contains conclusions.

## II. CONFLICTING ATC CLEARANCES

This section defines the "clearance conflicts" that are to be detected by the CATC system. Our definition roughly follows the one in [9]. We consider four types of runway-related ATC clearances: *line up (LUP), cross (CRS), take-off (TOF)* and *land (LND)*. Based on these four, we define one type of conflict for every unordered pair of clearance types:

- *LUP/LUP*: two aircraft are cleared to line up from opposing runway entries on the same end of a runway; or: two aircraft are cleared to line up on opposite ends of the same runway (*); or: two aircraft are cleared to line up on the same or adjacent runway entries on the same runway, and multiple line-up is not authorised (*).
- *LUP/CRS*: one aircraft is cleared to line up and another mobile is cleared to cross the same runway from an opposing runway entry.
- *LUP/TOF*: one aircraft is cleared to line up and another is cleared to take off on the same runway, and the runway entry of the aircraft lining up is in front of the position of the aircraft taking off.
- *LUP/LND*: one aircraft is cleared to line up and another is cleared to land on the same runway, and the runway entry of the aircraft lining up is in front of the position of the landing aircraft, and the landing aircraft is not expected to vacate the runway before the line up point.
- *CRS/CRS*: two mobiles are cleared to cross the same runway from opposing runway entries.
- *CRS/TOF*: one mobile is cleared to cross and another is cleared to take off on the same runway, and the runway entry of the crossing mobile is in front of the position of the aircraft taking off.
- *CRS/LND*: one mobile is cleared to cross and another is cleared to land on the same runway, and the runway entry of the crossing mobile is in front of the position of the landing aircraft, and the landing aircraft is not expected to vacate the runway before the crossing point.
- *TOF/TOF*: two aircraft are cleared to take off on the same runway.
- *TOF/LND*: one aircraft is cleared to take off and another is cleared to land on the same runway.
- *LND/LND*: two aircraft are cleared to land on the same runway.

The meaning of (*) is explained in Section IV.

Above and elsewhere in this paper, the term "runway" refers to a physical runway surface on the airport, for which there may be several runway thresholds. This implies that the above definitions of the LUP/TOF, LUP/LND, TOF/TOF and LND/LND conflicts also apply to movements that use opposite thresholds of the same runway. For example, a situation where two aircraft are cleared to land on thresholds "05" and "23" of runway "05/23" constitutes a LND/LND conflict.

The definition above only takes into account conflict situations on a single runway. We extend it to *intersecting* runways by stipulating that TOF/TOF, TOF/LND and LND/LND conflicts also occur if the two respective aircraft are cleared not on the same runway but on intersecting runways and if their trajectories are converging, i.e., if both will move towards the intersection, not away from it. We do not consider converging but non-intersecting runways here.

In principle, a detection mechanism for clearance conflicts can be implemented directly along the lines of the above definition. However, in this paper we argue that it is advantageous to instead base the detection on *ground routes*.

## III. GROUND ROUTING

Safety support mechanisms such as CATC constitute one functional area in the context of A-SMGCS or tower ATC systems. Another functional area is *ground routing*, where the system maintains the routes that mobiles are planned to follow on the airport's network of runways, taxiways, apron taxilanes and parking stands. System knowledge about these routes is an essential enabler for several other system functionalities. For example, possible functionalities that depend on route information are conformance monitoring, which warns if a mobile deviates from its assigned route; automatic switching of airport lighting to guide aircraft from their parking stand to the runway or vice versa; or the computation of accurate individual taxi time estimates to be used e.g. by a departure management planning system.

In this paper, we propose to use routes for CATC as well. To this end, we assume in Section IV that there is a route for every mobile that may receive a runway-related ATC clearance, and that this route matches ATCO intent. It is not relevant how the routes in the system actually come to be. Conceivable possibilities include manual input by the ATCO; a range of automatic approaches, from table-based lookup via simple shortest-path search algorithms to more complex time-based optimisation algorithms; or combinations thereof. See e.g. [12] for a literature survey, or [13] for an example of recent research in the area.

We define the notion of a *route* based on a *segmentation* of the airport map, i.e., a partitioning of the movement area into non-overlapping polygons called *segments*, such that every segment can be considered atomic for the purposes of routing. An example is shown in Figure 1.

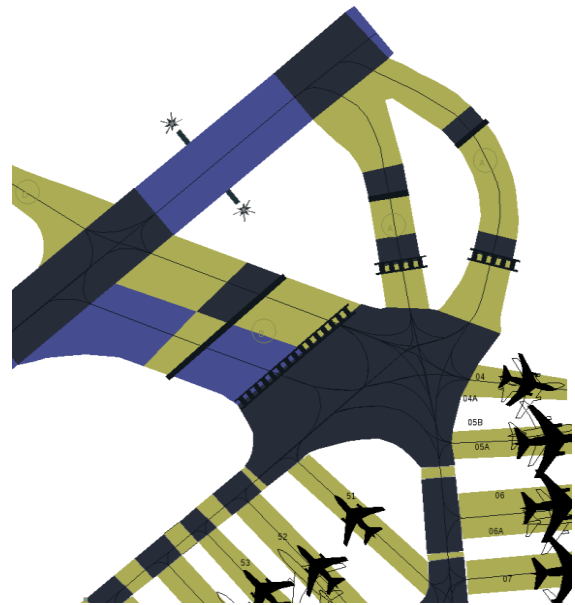

Figure 1: Airport segmentation; colors serve to highlight segment boundaries

Figure 1 depicts a possible segmentation of the northeastern corner of Hamburg airport. We see the northeastern end of runway 05/23, with four runway entry taxiways leading onto the runway from its south-east and one from its north-west. In the bottom right part of the figure some parking stands are visible. The segmentation is chosen such that every intersection between runways, taxiways and/or parking stands has its own segment, and such that there is a segment boundary at every holding line.

Given such a segmentation of the airport, we can understand a route abstractly as a *sequence of segments*. An example for a route in this sense is visualised in Figure 2.

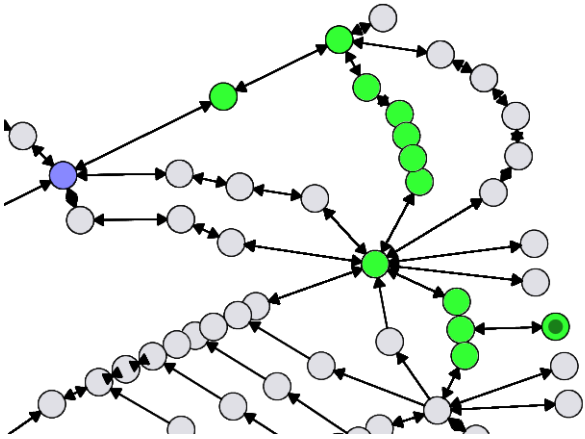

Figure 2: Ground route, consisting of those airport segments (shown as circles) that are colored green or blue

The route depicted in Figure 2 leads an aircraft from its current position on a parking stand (the segment on the right marked with a dot inside the green circle) first through a pushback and then to runway 05/23, and finally down that runway, representing take-off from logical runway 23.

Given polygonal paths that model the centerlines of taxiways and runways, one can derive from a route (i.e., from a sequence of segments) a smooth polygonal path corresponding to the route, by composing the centerline parts that belong to the route's segments. Such a polygonal path can e.g. be used to visualise the route in a human machine interface (HMI). An example is shown in Figure 3. Our CATC detection approach uses only the abstract route, not the corresponding polygonal path.

We assume that the route of an arriving aircraft initially starts at the runway threshold and ends at the parking stand, and that conversely, the route of a departing aircraft initially starts at its parking stand and ends at the far end of the take-off runway. Additionally, we assume that the routing subsystem continuously updates routes based on surveillance information, such that for a mobile that is proceeding to move along its route, the route always starts in the segment corresponding to its current position. A consequence of this is that the availability of accurate surveillance data is a necessary prerequisite for the availability of accurate routes, and that using routes for further purposes – such as CATC – implies using surveillance data.

When a mobile has both a clearance such as LUP, CRS, TOF or LND and a route, we can distinguish between an initial *cleared* part of the route that the mobile already has been allowed to execute, and the subsequent *planned* part that it is not yet permitted to execute. If a segment belongs to the cleared part of a mobile's route, we say that the mobile is *cleared for the segment*.

Roughly, for a LUP clearance, the cleared part of a route consists of all route segments up to and onto the take-off runway, but not yet down the runway. For a CRS clearance, it includes the route segments onto and beyond the first runway that the route crosses. For a TOF clearance, the cleared part is the entire route up to the end of the take-off runway, and for a LND clearance, it consists of all segments of the landing runway up to the planned runway exit. For example, the aircraft of Figures 2 and 3 currently has a LUP clearance. The cleared part of the route is shown in green in Figure 2, and as a solid line in Figure 3. The part of the route that the aircraft is not yet cleared for is shown in blue in Figure 2, and as a dashed line in Figure 3.

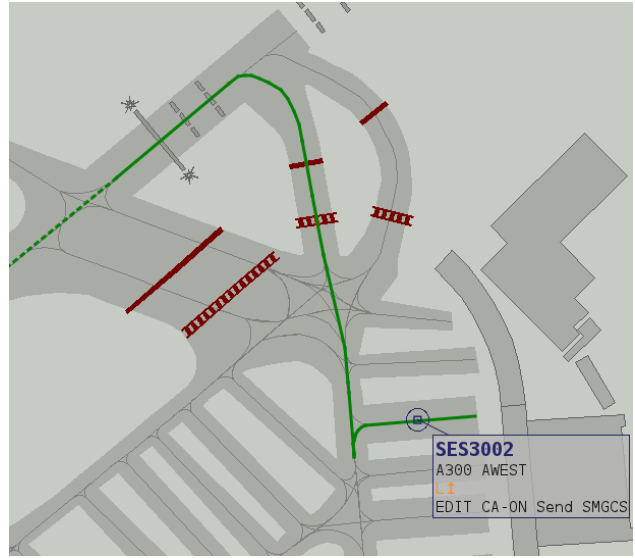

Figure 3: Polygonal path corresponding to the route of Figure 2, displayed in a tower controller HMI

IV. ROUTE-BASED CATC DETECTION

Looking again at the description of CATC in Section II, but ignoring the two cases marked with (*) for the moment, we can observe that the various conflict type definitions all follow one of two basic patterns. The first pattern is that for a mobile with a TOF or LND clearance, there is a conflict with any other mobile whose clearance permits it to use the runway ahead of the taking off or landing aircraft. It does not matter what clearance the other mobile has, or what direction it is moving in.

The second pattern underlies those conflicts that involve neither a TOF nor a LND clearance, i.e., conflicts only between LUP and/or CRS. For these "slow" clearances, there is a conflict only if the clearances allow the mobiles to meet on the runway while moving in different directions. Several overlapping movements in the same direction do not constitute a conflict; think e.g. of multiple aircraft consecutively crossing a runway from the same runway entry.

Using the terminology of Section III, these patterns can together be phrased as: there is a conflict between two mobiles if there is a runway segment for which both are cleared, unless (1) both clearances are either LUP or CRS and (2) both mobiles approach the runway segment in question from the same direction. We say that two mobiles "ap-proach a segment from the same direction" if the segment preceding the one in question is the same in both mobiles' routes.

Our approach is to use this unified, route-based formulation of CATC as the conflict detection logic. The pseudocode below determines the conflicts that a given mobile A is involved in:

*for every runway segment S on A's cleared route part:*
*for every mobile B ≠ A also cleared for S:*
*if not (A and B both have LUP or CRS clearance*
*and both approach S from the same direction):*
*report conflict between A and B*

The type of a thus detected conflict is determined by the clearances of the involved mobiles A and B. For example, if one of them has a LND clearance and the other a LUP clearance, then the conflict is a LUP/LND conflict.

The following examples are intended to illustrate that the above detection logic matches the conflict definitions in Section II, without the cases marked with (*). A first example situation is depicted in Figure 4.

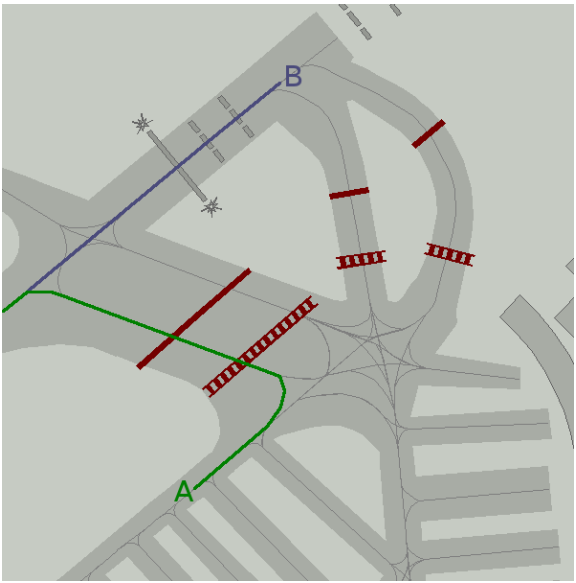

Figure 4: Example situation with two aircraft A and B and associated ground routes; the routes meet at the intersection and overlap in all segments further down the runway

Whether there is a conflict in the situation of Figure 4 depends on the clearances of A and B. If, for example, A and B both have LUP clearances, then there is no conflict, because there is no runway segment that both are cleared for. If, on the other hand, A has LUP clearance and B has TOF or LND clearance, then there is a LUP/TOF or a LUP/LND conflict, respectively, because there are runway segments that both aircraft are cleared for. In the terminology of Section II, the conflict occurs because the runway entry of A is in front of the position of B. If B moves down the runway beyond A's runway entry, then eventually the routes stop overlapping and the situation becomes conflict-free.

Figure 5 shows a second example situation. After joining paths on a taxiway, aircraft A and B approach their single shared runway segment from the same direction. If A has TOF clearance and B has CRS clearance, then there is a CRS/TOF conflict. But if A instead has LUP clearance, then there is no LUP/CRS conflict. Such a conflict would occur only if B were moving along its route in the opposite direction, because then, the two aircraft would be approaching their shared runway segment from different directions.

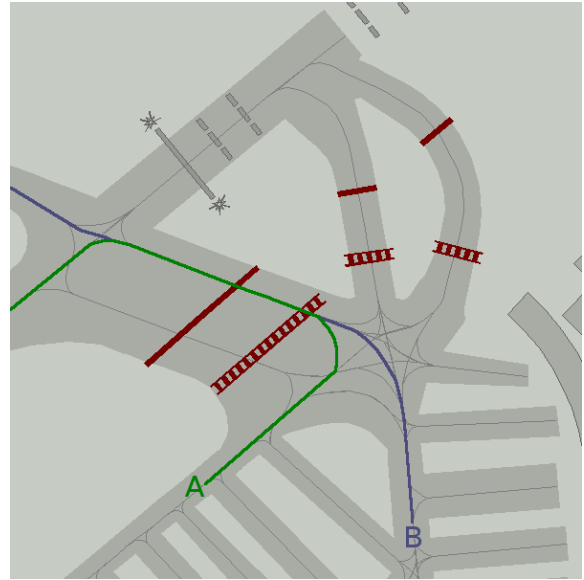

Figure 5: Second example situation; the routes of A and B overlap in exactly one runway segment

One of the advantages of the route-based approach is that it immediately works for intersecting runways, too. Imagine, for example, a pair of intersecting runways. Two aircraft with TOF or LND clearances for the two runways create a TOF/TOF, TOF/LND or LND/LND conflict, as desired, as they are both cleared to use the intersection segment shared by both runways. And, as desired, the conflict is resolved when one of them passes the intersection, because then the routes do not overlap anymore. A screenshot of a LND/LND conflict on intersecting runways is shown in Figure 6.

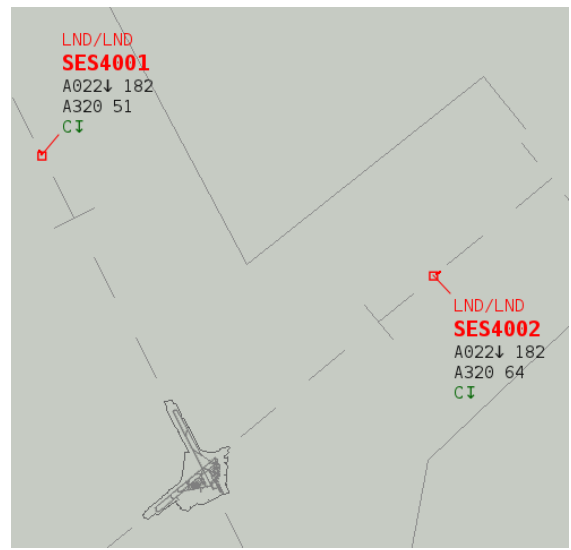

Figure 6: LND/LND conflict on intersecting runways, in an HMI

The definitions in Section II specify that LUP/LND and CRS/LND conflicts only occur if the landing aircraft is not "expected" to vacate the runway before the runway entry used by the second aircraft. They do not further specify what constitutes such an "expectation". The matter is tricky, because it is possible that aircraft miss their assigned runway exit, e.g. due to bad braking conditions. In the above ap-proach, we consider the routes to be the manifestations of the expectations that are to be used for CATC. The routing mechanism might ensure that by default, all landing aircraft have routes where they vacate at the very end of the runway, and if an ATCO expects the use of an earlier exit, he or she may change the route accordingly. Alternatively, one could extend the definition and implementation of CATC with some distinction between uncertain expectations (say, the runway exit planned to be used by a still-airborne aircraft) and expectations with a higher degree of confidence (such as the runway exit planned to be used by an aircraft that has already touched down and decelerated, or perhaps by an aircraft with a reliable "brake to vacate" system).

We have so far ignored the cases of the LUP/LUP definition that are marked with (*) in Section II, i.e., the cases where two aircraft line up on opposite runway ends, or where they line up behind each other and where this is disallowed by local rules. These conflicts follow a different idea than the others: in all other conflicts, one can imagine situations where, if the two aircraft do what their clearances allow them to do, this would lead to a collision on the runway. In contrast, the worry behind the cases marked with (*) has more to do with the fact that LUP is usually followed by TOF, and that the TOF clearance in these cases would be problematic. Supporting these cases in our CATC detection logic requires adding a special case, which is nevertheless easy to express using routes: if A's clearance is LUP, then we do not only check the cleared part of its route, but also the not-yet-cleared runway segments that belong to its take-off, and we report a conflict if there is another aircraft that holds a LUP clearance for any of these runway segments.

## V. CONDITIONAL CLEARANCES

A special form of clearances on airports are *conditional clearances*, i.e., clearances that become effective only when a certain condition is satisfied. Most frequently, the clearance in question is a LUP or CRS clearance, and the condition is that a landing or taking off aircraft must have passed the runway entry point to be used for lining up or crossing, respectively. Conditional clearances may be used by ATCOs to inform pilots about clearances "in advance", giving them the permission to start the corresponding movement at a later point in time, namely when the pilot is certain that the specified other aircraft has passed his own.

In CATC detection, a conditional clearance must be treated differently from a regular clearance. For example, a conditional LUP clearance to line up behind a landing aircraft should not create a LUP/LND conflict between the two aircraft. In a CATC sense, a conditional clearance is not really a clearance at all until the moment when the attached condition is fulfilled. At this point in time, it turns into a regular, unconditional clearance.

The CATC logic can be used to automatically determine this moment, and to then automatically remove the condition in the flight strip. This should happen as soon as doing so does not create a clearance conflict with the "condition" aircraft. For example, the system can thus automatically update a conditional LUP clearance to a proper LUP clearance at the precise moment when the landing aircraft named in the condition has passed the runway entry point, i.e., at the first moment when this update does not create a LUP/LND conflict between the two aircraft anymore.

## VI. HUMAN MACHINE INTERFACE

In order to inform ATCOs about detected CATC conflicts, warnings may be displayed in sensor data processing and/or electronic flight strips HMIs. We have already seen an example for the former in Figure 6. An example for a warning in an electronic flight strip is shown in Figure 7 below, where a CRS/TOF warning is attached to the strip of flight "DLH017". The same warning would also be attached to the strip of the second mobile involved in the conflict.

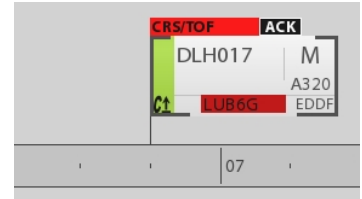

Figure 7: Clearance conflict warning in flight strip

Warnings may be displayed on all the different controller working positions (e.g. Ground Delivery, Apron Controller, Taxi Controller, Runway Controller, Supervisor Controller, etc.). It may be possible to turn on and off the presentation individually on every working position.

Conditional clearances should be displayed such that the difference to regular clearances is clearly visible. One attempt to do this is shown in Figure 8, where the conditional clearance of flight "FDX111" to line-up behind flight "SAS638" is indicated by the callsign of "SAS638" being shown in red letters above the label of "FDX111".

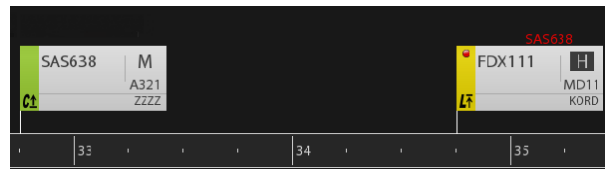

Figure 8: Conditional clearance in flight strip

The CATC warnings shown in Figures 6 and 7 are about *current* conflicts, i.e., conflicts between currently active clearances. These warnings typically appear as soon as an ATCO enters a clearance that is in conflict with another one. It is also possible to display warnings about *future* conflicts, *before* the corresponding clearance gets entered into the system. Such predictive warnings could help ATCOs to avoid a mistake before it even happens.

For predictive warnings, the system can internally use the standard CATC logic to check whether certain clearances would lead to conflicts or not. This may be done on-demand – perhaps when a menu displaying possible clearances gets opened in the HMI – or periodically for all clearances that can possibly or likely be issued in the current situation.

An example for a predictive indication is shown in Figure 9 below, where a green "probe light" is displayed on the upper right corner of the button that ATCOs use for entering the "expected" next clearance, i.e., the one that will typically be next in the workflow on the airport. This tells the ATCO that giving this clearance at the current moment in time will not lead to a clearance conflict. If giving the next expected clearance *would* lead to a conflict, then the strip is shown with a red probe light on the clearance entry button instead of a green one; an example for this can be seen in Figure 8 above, where the label of "FDX111" has a red probe light because giving it the next expected clearance (i.e., TOF) would create a TOF/TOF conflict with "SAS638".

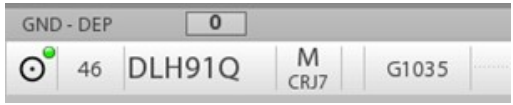

Figure 9: Flight strip with green "probe light", indicating that giving the next clearance now will not create a clearance conflict

## VII. Trajectory-Based CATC Detection

The definitions in Section II aim at warning about situations where a pair of clearances could lead to a safety problem, such as a violation of runway separation. They are by nature an approximation: the presence of a conflict does not imply that there will necessarily be a problem. In particular, it is an approximation in that the dimension of time is abstracted away completely. Consider e.g. a CRS/LND conflict: if the crossing mobile is already about to vacate the runway, and if the landing aircraft is still minutes away from the airport, the situation may be safe and acceptable according to the rules of runway separation. If in practice there are too many such warnings about situations that ATCOs consider to actually be safe, then this could severely limit the overall practical usefulness of the safety system.

From our perspective, it remains to be studied how adequate the degree of approximation of the current CATC concept as described in Section II and in [9] is in practice. It is clear already that a significant amount of "fine-tuning" will be needed to adjust the conflict definitions to the local rules or customs of an individual country or airport. Beyond that, we feel that it may be necessary to more generally extend the concept of CATC to take into account the dimension of *time*. The route-based approach appears well suited for such an extension: based on predicting the times at which mobiles are expected to reach the segments of their routes, one could extend the two-dimensional routes to two-plus-one-dimensional trajectories that include time in addition to geographical information. Based on these trajectories and on a precise model of the applicable set of rules for runway separation (which can be somewhat complicated), one could create a system that warns about a pair of clearances only if the pair is expected to lead to a violation of the required runway separation, or if – using a probabilistic model – it does not appear sufficiently likely that runway separation will be maintained throughout.

The vertical dimension may be of interest, too: for example, the notion of CATC may be extended to conflicts for converging but non-intersecting runways, where the problem is in the surrounding airspace rather than on the runways themselves. Thus, ultimately, a CATC system may be based on 4D trajectories, mainly on their ground parts but also on the parts in the vicinity of the airport. Such a CATC system is then closely related to *medium-term conflict detection (MTCD)* systems [14] in en-route ATC, which – based on clearances and predicted trajectories – warn about en-route aircraft that might get too close in the future.

## VIII. Conclusion

We have presented an approach for detecting conflicting ATC clearances on airports based on ground routes. Using ground routes – which are a central enabler for many other functions in future airport ATC systems, too – allows for a rather simple core safety logic that immediately also works for less obvious cases like intersecting runways. A route-based approach is also naturally suited for a future extension to a trajectory-based conflict detection, which would be more complex but could also be more accurate in avoiding superfluous warnings.

The route-based approach has been implemented prototypically as an extension of the *DFS PHOENIX* sensor data processing system [15] and the *DFS SHOWTIME* electronic flight strips system [16], and the prototype has been used in a SESAR validation trial about conflicting ATC clearances conducted at Hamburg airport [17].

## Acknowledgements

We would like to thank department SH/P and especially Hendrik Glaab, Ruben Nunez, Marco Siciliano and Stephan Unger for programming support; Klaus Pourvoyeur for help with the PHOENIX tracker; Marc Bonnier and Roger Lane (EUROCONTROL) for useful discussions; and Heribert Lafferton (DFS) as well as Marcus Biella, Marcus Helms and Karsten Straube (DLR) for organising the SESAR validation exercise that allowed us to test our ideas. The work behind this paper was conducted in the framework of SESAR, which is co-financed by the European Community and by EUROCONTROL.